\documentstyle[12pt]{article}              

\title{{\hfill{\small{BGU PH-97/02}}}\protect\\ 
Inclusive spectra of hadrons
 created by color tube fission 2. Inclusive spectra of primary hadrons}
\author{E.V.Gedalin\thanks{E-mail address: gedal@bgumail.bgu.ac.il}}
\date{Department of Physics \\
 Ben-Gurion University of the Negev\\
Beer-Sheva, 84105,  Israel}
\begin{document}
\maketitle
\newpage
\begin{abstract}
The primary inclusive spectra and correlation functions of particles 
created by color tube fission are considered. Using the previously 
obtained expression for probability  of the tube breaking in $n$ points 
we have calculated the one and two particle inclusive spectra of 
tube pieces as well as pseudoscalar and vector mesons in plateau area. It 
is shown that the plateau height of the one particle inclusive spectrum  is 
determined by the flavor quark composition and spin of hadron. Small 
oscillations of the tube surface give only small correction to the main 
term. The correlation functions  of fixed 
particles have the form of a product of the universal function that depends 
only on the particle rapidity difference and thescale factor dependent on 
the spin and flavor quark composition of hadron.  
\end{abstract}
\pagebreak

\section{Introduction}
\bigskip
\ \ \ In the preceding article \cite{Ge97} we have calculated the probability 
of the tube fission in  one and many points. Here we apply the obtained 
expressions  to the calculation of the inclusive spectra (IS) of 
primary hadrons produced in soft processes. Our 
approach to the tube fragmentation into hadrons is close to Artru- 
Menessier model \cite{Ar83} (as well as other models \cite{An83}-\cite{Gn} 
based on Artru - Menessier approach). In our model \cite{Aggk}-\cite{Ge93}
at each stage of hadronization process the tube splits 
into pieces with arbitrary masses 
until the distance between the $K$ and $\bar{K}$ walls of piece becomes of 
the order of thickness of the wall. This causes growth of the kink mass and 
rapid decrease  of the probability of the tube fission and therefore the 
hadronization process stops. 
The produced pieces are of the length 
$\stackrel{>}{\sim} 2M / \epsilon^2$, where $M$ and $\epsilon^2$ are the 
kink mass and tube tension respectively.

The piece IS are proportional to the probability of the tube breaking in 
two (and more) points that forms the  piece with the light cone momentum 
$p_+$ and $p_-$, mass $m$ and transverse momentum $k$. 
The usual mass shell condition 
$p_+p_- = k^2 + m^2$ eliminates $p_-$ and we must calculate probability 
of the tube breaking in two or/and more points with the conditions for 
the light cone coordinates of piece edges 
\begin{eqnarray}
	 && u_1 - u_2 = p_+/\epsilon \nonumber \\
	 && v_2 - v_1 = p_-/\epsilon = (m^2 + k^2)/p_+\epsilon
	 \label{dif}
\end{eqnarray}
etc. The obtained probabilities depend on $p_+$ as well as on the piece 
transverse mass squared, i.e. the breaking mechanism fully describes the 
IS of pieces (see the ref. \cite{An83}-\cite{We93}).

Since far we are interested only in the $p_+$ -dependence  of IS the $k^2$ 
dependence must be integrated and we obtain the condition
\begin{equation}
	v_2 - v_1 \geq m^2/p_+\epsilon
\end{equation}   
thatgreately simplifies the calculations. Usually the rhs of this 
condition is taken equal to zero, neglecting all mass corrections. 
However, the mass of the piece  plays an important role having influence on 
the value  of the corresponding probability. We shell return to the 
mass corrections later.

When the mass corrections are neglected, the piece mass is not well 
defined that the pieces can be attributed to stable hadrons (pions 
and kaons) as well as hadron resonances. In other words, the piece 
with the flavor quantum numbers  $i$ and $j$ of a kink and antikink has the 
probability $C(i,j; S)$ to be in state $S$, which is the stable hadron 
or resonance. The advantage of this concept is that all corrections 
necessary for including an additional resonances and final state 
interactions, can be implemented quite naturally (see for example refs. 
\cite{An83}-\cite{We93}).

We assume that the primary mesons are pseudoscalar and vector particles,  
neglecting temporarily the diquark  and other type of $K,{\bar{K}}$  
production that can be introduced in the same way (for a more complete 
discussion we refer to Ref.\cite{An83} -\cite{We93})

The article is organized as follows. First, we calculate the IS of the 
pieces that have fixed quark quantum numbers on the edges. The 
primary hadron IS then are built from the IS of pieces, using the quark 
flavor composition of hadrons. We take into account only $u, d$ and 
$s$ quarks. The probabilities $w_u$ and $w_d$ of the $u$ and $d$ quark 
production are taken the same $w_u = w_d = w$ and the (rare) strange quark 
creation probability $w_s$ is assumed to be sufficiently smal. We have 
$2w + w_s = 1$. In the next section we calculate the one particle 
inclusive spectra (OPIS) of pieces . The two particle inclusive spectra 
(TPIS) of pieces and corresponding correlation functions (CF) are 
considered in Sec.3. The pseudoscalar and vector meson primary IS are 
calculated in Sec.4.In Conclusion we summarize and discuss the obtained 
results.
\section{The OPIS of pieces}

\bigskip

Here we calculate inclusive spectra of primary hadrons created 
by fission of a tube with small surface oscillations. We limit ourselves 
with particles  with $p_+$ small in comparison with the momentum of initial 
tube $p_{0+}$ (but $p_+>>m$). We restrict ourselves with IS integrated over 
transverse momenta and neglect mass corrections.

First we consider the one-particle IS (OPIS). Let us begin with the 
case of one type of $(K, \bar{K})$, that is, where we detect the 
primary hadrons of any quantum numbers. The 
procedure of one-particle IS calculation is the following. We must 
integrate over the probability $dP(1, 2)$ of two adjacent fission at 
$(u_1, v_1)$ and $(u_2, v_2)$, that form the piece with $p_+$, over all 
positions of fission the space-time points. 
From eq. (70) of ref. \cite{Ge97} we have
\begin{equation}
dP(1,2;p_+) = dw(1)dw(2) \exp [-W(\tilde{S})] \vartheta (u_1 - u_2)
\vartheta(v_2 - v_1) \delta (u_1 - u_2 - \tilde{p})\label{prob8}
\end{equation}
where the integration region $\tilde{S}$ is the rectangle area defined by 
lines $u=0, u=u_1, v=0, v=v_2$. The $\vartheta$ - functions in expression 
(\ref{prob8}) reflect the ordering of fission points and $\delta$ - function 
accounts for the difference condition (\ref{dif}).

Now the OPIS is given by 
\begin{equation} 
f_1(\tilde{p}) =  \int dP(1,2;p_+),
\end{equation}
where the integration is over the range $(0, \tilde{p_0})$ 
for each variable .\footnote[1]{Only a small ($\sim 1/\tilde{p}_0$) 
contribution arise from one breaking point diagram when one wall of 
produced piece is the wall of initial tube and only second wall is 
produced by additional $K \tilde{K}$ pair.  We will systematically 
neglect such contributions to IS in the plateau area}

As we have seen above the small-mass oscillations produce  corrections to 
the fission probability of long tube that are proportional to the oscillation 
amplitude squared (see Ref.\cite{Ge97}), namely   
\begin{equation}
dw(\tau, \eta) = dw_0 \exp (a^2 D(\kappa s)).
\end{equation}
where 
\begin{equation}
dw_0 = (\epsilon^2/2\pi)\exp (-\pi M^2/\epsilon^2)d\eta d\tau = 
w_0 d\eta d\tau	
\end{equation}
is the fission probability without oscillations, $D = (1/\epsilon^2)D_{(n)}(s)$ 
 depends on the initial shape of the color tube, 
$\kappa$ is the (dimensionless) mass of the small mass surface oscillations,
 $s_2 = 
\tau^2 - \eta^2 = uv$. It is important that $D$ depends only on the product 
$uv$ and not on $u$ and $v$ separately.

 Since we consider small oscillations with $a^2 \ll 1$ we expand 
 all expressions in a power series in $a^2$ end keep terms up to $a^2$ 
 order. After some calculations we have
 \begin{equation}
 f_1 = (1/\tilde{p})(1 -  a^2 F(q)),
 \end{equation} 
where $q = 2\kappa^2/w_0$ and
\begin{equation}
F(q) = \sum_{n=1}^{\infty} q^n Z_n \frac{n!n}{n+1},
\end{equation}
$Z_n$ are the coefficients of power expansion of function $D = \sum q^nZ_n$.

 We see that small oscillations do not affect the spectrum but change 
the height of the plateau \cite{Gg90b}. Obviously this change 
depends on the probability the of tube fission.

To proceed to the case of many type of kinks we note that the probability 
of the tube breaking at the point $(u,v)$ due to the creation of the
$(K, \bar{K})$ pair of type $l$, can be written in the form 
\begin{equation}
dw(u,v; l) = w_le^{-a^2 Z_l(ks)}dudv
\end{equation}
and
\begin{eqnarray}
&&W(\tilde{S}) = \frac{1}{2}\int_{\tilde{S}} du' dv' w_{tot}\nonumber \\
&&= \frac{1}{2}\int_{\tilde{S}} du' dv'\sum_l w_l (1 + a^2\sum_n 
Z_n(l)(ku'v')^n)\label{prob6}
\end{eqnarray}
where $w_l$ is the partial fission probability without oscillations and 
$Z_n(l)$ are the coefficients of the power expansion of function $Z_l(ks)$.

Repeating the procedure of OPIS calculation we obtain for particle with 
$K$ oftype $j$ and $\bar{K}$ of type $l$ on the edges the 
expression \cite{Gg90b}
\begin{equation}
f_1(j,l;\tilde{p}) = \frac{r_jr_l}{\tilde{p}} (1 -a^2 F(j,l; q))
\end{equation}
where $r_l = w_l/w_{tot}$ is the relative weight of the kink of flavor $i$ 
and $w_l$ and $w_{tot} = \sum_l w_l$ are the partial and total breaking 
probabilities without oscillations respectively and
\begin{equation}
F(j,l;q) = \sum_n \Gamma(n+1) q^n[Z_n(j) + Z_n(l) - 
\sum_k r_l\frac{n+2}{n+1}Z_n(l)]
\end{equation}

As in the previous case the momentum spectrum is not affected. 
However the height of the plateau is defined by now the product 
$r_jr_l$ (with small corrections $\sim a^2$) and is  
$(j,l)$ dependent. In the case when on both edges of the piece are 
dominated kinks ( i.e. the kinks of small mass with large creation 
probability $w_l\approx w_{tot}$) we obtain the plateau height close to 
unity as in the case of one type kink. The (strange) kinks with 
$r_l \ll 1$  contribute only into $\sim a^2$ corrections. When one 
(or both) kink is strange  the height of the plateau is small. However the 
$a^2$ - order corrections contain contributions of dominant as well as 
strange kinks.

\section{The TPIS and CF of the pieces}

Now we pass to the two particle inclusive spectra (TPIS) and particle 
correlations. For TPIS we have two distinct contributions.  The first one 
$f_2^{(1)}$ is from the two pieces produced in three adjacent fission 
points $(u_1, v_1), (u_2,v_2)$ and $(u_3,v_3)$ ($(1), (2)$ and $(3)$), 
having one of the edges created in the same fission at point $(2)$ 
while there are no fission between points $(1)$ and $(2)$ as 
well as between $(2)$ and $(3)$. These pieces are the primary particles 
of adjacent ranges in Field-Feynman terminology \cite{Ff}.

The second contribution $f_2^{(2)}$ is from two by two adjacent 
fission $[(u_1, v_1), (u_2, v_2)]$ and $[(u_3, v_3), (u_4, v_4)]$ 
and there is no fission between the points $(1)$ and $(2)$ 
as well as between $(3)$ and $(4)$.

Again first consider the one flavor case \cite{Gg90b}. 
The first contribution has the form \cite{Gg90b}.
\begin{eqnarray}
&&f_2^{(1)}(\tilde{p}_1, \tilde{p}_2)\nonumber \\ 
&&= \int [\prod_{i=1}^{3}dw(i)] \exp[-W(u_1,v_2)-W(u_2, v_3)
+W(u_2,v_2)]\vartheta (u_1-u_2)\\
&&\vartheta (u_2-u_3)\vartheta (u_3)\vartheta (v_2-v_1)
\vartheta (v_3-v_2)\vartheta(v_1) 
\delta (u_1-u_2-\tilde{p}_1)\delta (u_2-u_3-\tilde{p}_2)\nonumber \\
&&+ (\tilde{p}_1 \Leftrightarrow \tilde{p}_2)\nonumber
\end{eqnarray}
where we have denoted by $W(x, y)$ the $W(S)$ defined by expression 
(\ref{prob6}) for the rectangle area limited by 
lines $u=0, u=x, v=0, v=y$  and $(\tilde{p}_1 \Leftrightarrow \tilde{p}_2)$ 
 is the $\tilde{p}_1$ and $\tilde{p}_2$ interchanged term. The 
first term describes the case when tube $(1)-(2)$ has  
$u_1 - u_2 = \tilde{p}_1$ and tube $(2)-(3)$ has  $u_2 - u_3 = 
\tilde{p}_2$ and the second term describes the situation when $u_1 - 
u_2 = \tilde{p}_2$ and $u_2 - u_3 =\tilde{p}_1$.

The calculation is straightforward but somewhat cumbersome and we obtain 
\begin{equation}
f_2^{(1)} = \frac{1}{\tilde{p}_1\tilde{p}_2}[1 
+\varphi (\tilde{p}_1, \tilde{p}_2)][1-a^2 F(q)]
 + \frac{a^2}{\tilde{p}_1 \tilde{p}_2} \varphi_1 (\tilde{p}_1/\tilde{p}_2, 
 q),
\end{equation}
 where $ \frac{1}{\tilde{p}_1\tilde{p}_2}[1 + 
 \varphi(\tilde{p}_1, \tilde{p}_2)]$ is the spectrum of the two pieces 
 produced in the adjacent fission without oscillations and
 \begin{eqnarray}
 	 & & \varphi(\tilde{p}_1, \tilde{p}_2) =
 	  \varphi(\tilde{p}_1 / \tilde{p}_2) 
 	\label{phig} \\
 	 & & \varphi(x) = x\ln (1 + \frac{1}{x}) + \frac{1}{x} \ln (1 + x) -1
 	\label{phi} 
 \end{eqnarray}
 The function $\varphi_1$ depends only on rapidity difference $y = \ln 
 (\tilde{p}_1/\tilde{p}_2)$ and his explicit form is closely related to 
 the form of $Z(x)$.

 The second contribution $f_2^{(2)}$ can be written in the form
\begin{eqnarray}
&&f_2^{(2)} = \nonumber \\
&& \int [\prod_{i=1}^{4}dw(i)] \exp[-W(u_3,v_4) -W(u_1, v_2)
 + W(u_3,v_2)]\\
&&\vartheta (u_1 - u_2)\vartheta (u_2 - u_3)\vartheta(u_3 - u_4)\vartheta 
(u_4)\vartheta (v_2 - v_1)\vartheta (v_3 - v_2)\nonumber \\
&&\vartheta (v_4 - v_3)\vartheta (v_1)\delta (u_1 - u_2 - \tilde{p}_1)
\delta (u_3 - u_4 -\tilde{p}_2)\nonumber\\
&& + (\tilde{p}_1 \Leftrightarrow \tilde{p}_2)\nonumber
\end{eqnarray}

After some calculation we obtain \cite{Gg90b}.
\begin{equation}
f_2^{(2)} = f_2^{(1)} + \frac{1}{\tilde{p}_1\tilde{p}_2}(1 - a^2[2F(q) + 
R(q,y)])
\end{equation}
where 
\begin{equation}
R(q,y) = \frac{e^y}{1+e^y}F(\frac{qe^y}{1+e^y})
 + \frac{1}{1+e^y}F(\frac{q}{1+e^y}) - F(q).
 \end{equation}
 and $y = ln (\tilde{p}_1/\tilde{p}_2)$ is the particle rapidity 
 difference.

 Thus $f_2$ has the form
 \begin{equation}
 f_2 = \frac {1}{\tilde{p}_1\tilde{p}_2}(1 - a^2[2F(q) -R(q,y)])
 \end{equation} 
and the correlation function (CF) is proportional to $a^2$ and depends 
only on $y$
\begin{eqnarray}
&&C_2(e^y) = f_2(\tilde{p}_1, \tilde{p}_2)/f_1(\tilde{p}_1)f_1(\tilde{p}_2) 
- 1 \nonumber \\
&&= a^2[e^y(1 + e^y)^{-1}F(qe^y (1 + e^y)^{-1})\\ 
&&+ (1 + e^y)^{-1}F(q(1 + e^y)^{-1}) - F(q)]\nonumber
\end{eqnarray}

As we can see $C_2$ in one flavor case the tube surface oscillations 
produce weak ($\sim a^2$) short range rapidity correlations with 
correlation length $\approx 1$. The CF $C_2$ is expressed by the same 
function $F(x)$ that $f_1$ is. At $y = 0$  $C_2(1, q) = a^2[F(q/2) - F(q)]$ 
and at large $|y|\gg1$ it vanishes exponentially
\begin{equation}
C_2(e^y) \sim - a^2e^{-|y|}[F(q) + qdF(q)/dq]
\end{equation}
Below we shall show that this correlation can be seen as small 
correlation between resonances.

Now we proceed to the multiflavor TPIS of particles $(i,k)$ and $(l,j)$ 
which internal quantum numbers defined by flavors of $K$ and 
$\bar{K}$ on the piece edges.

There are two different kinds of TPIS. The first one $f_{2s}(i,k: 
l,j;\tilde{p}_1, \tilde{p}_2)$ is 
symmetric in $\tilde{p}_1$ and $\tilde{p}_2$ as well as in $(i,k)$and 
$(l,j)$ pairs. The second type 
$f_{2n}(i,k;\tilde{p}_1; l,j,\tilde{p}_2)$ is 
symmetric only with respect to simultaneous transposition of 
$(i,k;\tilde{p}_1) \Leftrightarrow (l,j; \tilde{p}_2)$.

Let us consider first the $f_{2s}$. We begin with the three breaking point 
contribution $f_{2s}^{(1)}$. It is clear that  $f_{2s}^{(1)}$ vanishes 
unless one of the $i, j$ or /and $k,l$ pairs is the $(K, \bar{K})$ of 
same flavor. Therefore we have 
\begin{equation}
f_{2s}^{(1)} = r_i r_k(r_j\delta_{lk} + r_l\delta_{ij})
\frac{1}{\tilde{p}_1 \tilde{p}_2}[\varphi (\tilde{p}_1, \tilde{p}_2) - 
a^2\Phi_{2s}^{(1)}(i,k; lj; \tilde{p}_1, \tilde{p}_2)]\label{efone}
\end{equation}
where the terms proportional to  $\sim r_i r_k r_j \delta_{kl}$ 
and  $\sim r_i r_k r_l \delta_{ij}$ are the contributions 
when at a middle breaking point are 
created the $l = k$ or $i = j$  kink-antikink pairs respectively, and 
function $\Phi_{2s}^{(1)}$ describes the influence of tube surface 
oscillations .

The four breaking point  contribution $f_{2s}^{(2)}(i,k: l,j;
\tilde{p}_1, \tilde{p}_2)$ has a more simple form  
\begin{equation}
f_{2s}^{(2)} = r_i r_k r_j r_l
\frac{2}{\tilde{p}_1 \tilde{p}_2}[1 - \varphi (\tilde{p}_1, \tilde{p}_2) - 
a^2 \Phi_{2s}^{(2)}(i,k; lj; \tilde{p}_1, \tilde{p}_2)]	
\end{equation}
where again
the function $\Phi_{2s}^{(2)}$ describes the influence of small 
oscillations of tube surface.

Thus $f_{2s}$ has the form
\begin{equation}
f_{2s} = \frac{2}{\tilde{p}_1 \tilde{p}_2}[2 r_i r_k r_j r_l 
+ r_i r_k(r_j\delta_{lk} + r_l\delta_{ij} -2 r_l r_j)
\varphi (\tilde{p}_1, \tilde{p}_2) - 
a^2 \Phi_{2s}(i,k; lj; \tilde{p}_1, \tilde{p}_2)] \label{fsym}
\end{equation}
where 
\begin{equation}
\Phi_{2s} = 2 r_i r_k r_j r_l\Phi_{2s}^{(2)} + r_i r_k(r_j\delta_{lk} + 
r_l\delta_{ij})\Phi_{2s}^{(1)}
\end{equation}

From expression (\ref{fsym}) it follows that in the multiflavor case the two 
particle correlations are dependent on the internal quantum 
numbers of particles.

Let us consider the (symmetric) CF 
\begin{equation}
C_{2s} = 
\frac{f_{2s}(i,k; l,j;\tilde{p}_1, \tilde{p}_2)}{f_1(i,k;\tilde{p_1})
 f_1(l,j;\tilde{p_2}) + f_1(i,k;\tilde{p_2}) f_1(l,j;\tilde{p_1})} - 
 1.\label{csym} 
\end{equation}
When all four $K (i = k = l = j)$ are the same and dominant the CF has 
the form 
\begin{equation}
C_{2s} = \frac{1 - r}{r} \varphi (\tilde{p}_1, \tilde{p}_2) -
a^2 \Phi_{2s}(q,Y)
\end{equation}
The first term evidently describes the correlations without oscillations, 
the last term ($\sim a^2$) describes the influence of the surface 
oscillations.

It is clear that only at $r = 1$ all correlations are produced by the 
tube surface oscillations. Even if $r$ is very close to 1 it must 
the great influence of of first term must be expected. The oscillation 
role can be checked if we examine the $y$ dependence of $C_{2s}$ . Then 
the difference $C_{2s} - (1 - r)\varphi/r$ gives the measure of the 
influence of the oscillation term.

Let us consider now the case when there are $n \geq 2$ types of dominant 
kinks with equal $r_d$ and one  type of kink with $r_s < r_d$ (for 
instance the kinks that correspond to the $u$, $d$ and $s$ quarks 
respectively). Then $nr_d + r_s = 1$ and in order to obtain any information 
on the tube surface oscillations we must sum over all types of kinks (sum 
over full multiplet of quarks), i.e. to study correlations irrespective 
to the internal quantum numbers of pieces. From (\ref{fsym}) and 
(\ref{csym}) it easy to see that in this case $C_{2s} = a^2 R(q,y)$ just 
as in the one flavor case. Otherwise the $C_{2s}$ is strongly dominated 
by the nonoscillation term.
For correlation between particles with fixed quantum numbers  the 
influence of tube surface oscillations is very small. However, in contrast 
to the one flavor case now we have strong rapidity correlations. When 
only one pair of kinks is present (for definiteness $k$ and $l$) we have
\begin{equation}
C_{2s}(i,k,k,i;\tilde{p}_1, \tilde{p}_2) = 
(\frac{1}{2r_k} -1)\varphi (\tilde{p}_1, \tilde{p}_2)  
\end{equation}
The correlation become much more stronger for the rare ($K,\bar{K}$) 
pair ($s,\bar{s}$ quarks) creation in the middle point of three breaking 
point term. In this case  $r_k$ is small and TPIS is dominated  by the three 
breaking point contribution because the four breaking point term has an 
additional small $r_k$ factor. 

The same holds also for the case of two different pairs with flavors
$i = j$ and $k = l$ flavors where one pair of kinks is rare. The only 
difference from previous case is that we must observe the correlations 
between $K^0$ and $\bar{K}^0$ or $K^+$ and $\bar{K}^-$ ( or corresponding 
resonances).

Let us assume now that we detect only charged primary hadrons with same 
charges. Such particles cannot be created as primaries of adjacent ranges 
and none of the $i,j$ and $k,l$ pair may be of the same flavor. Then the 
contribution $f_2s^{(1)}$ vanishes. Therefore $f_{2s}$ has only a four 
breaking point contribution and we have 
\begin{equation}
C_{2s} = -  \varphi(\tilde{p}_1, \tilde{p}_2).\label{citwo}
\end{equation}
 It must be noted that here the sign of CF is opposite to the sign of 
 the previous case. 
 
 Finally we consider the $f_{2n}$ TPIS. The $(i,k,\tilde{p}_1) 
 \leftrightarrow (l,j,\tilde{p}_2)$ symmetry influences only on the three 
 breaking point term and instead of $f_{2s}^{(1)}$  we now have
\begin{eqnarray}
	 &&  f_{2n}(i,k;\tilde{p}_1; l,j,\tilde{p}_2) = \nonumber\\
   &&\frac{2}{\tilde{p}_1 \tilde{p}_2} r_i r_k[r_j\delta_{lk}\varphi_n (\tilde{p}_1, \tilde{p}_2)
 + r_l\delta_{ij} -2 r_l r_j) \\
&&\varphi_n (\tilde{p}_2, \tilde{p}_1)] \nonumber\\
	 && - O(a^2) \nonumber
\end{eqnarray}
 where 
 \begin{equation}
 \varphi_n (x, y) =\frac{y}{x} \ln[(x + y)/y] - \frac{y}{x + y}
 \end{equation}
 and $O(a^2)$ denotes the term proportional to $a^2$. Respectively for 
 $C_{2n}$ we obtain the expression
 \begin{equation}
 C_{2n} = f_{2n}(i,k;\tilde{p}_1; l,j,\tilde{p}_2)
 /(f_1(i,k;\tilde{p_1})f_1(l,j;\tilde{p_2}) -1
 \end{equation}
 and all conclusions of preceding discussion remain unchanged  also for 
 $C_{2n}$ with obvious substitution $\varphi \rightarrow \varphi_n$ in 
 formulas (\ref{fsym})-(\ref{citwo})
   
  The above consideration of IS and CF of pieces is sufficiently general 
  and can be used also for pieces with barion quantum numbers. The only 
  change is the substitution "rare quarks"  
 $\rightarrow$ "diquarks" with corresponding $r_{dq}$.

 \section{IS of primary hadrons}
 
 \bigskip
 
 The primary hadron IS can be constructed from IS of pieces in accordance 
 with hadron quark flavor composition and spin state. Here we restrict 
 ourselves by two nonets of mesons that are the pseudoscalar ($\pi, K, 
 \bar{K}, \eta$ ,$\eta'$)  and the vector $\rho, K*, \bar{K}*, \omega , 
 \varphi$ mesons. The other words we shell consider $P$ and $V$ meson 
 nonets built from $u$, $d$ and $s$ quarks and antiquarks and have 
 similar flavor structure. The oktet-nonet mixing for physical mesons is 
 taken into account  for $\eta$ and $\eta'$ and $\omega$ and $\Phi$ 
 mesons according to \cite{pd}-\cite{Gw}.
 
 Let us denote by $C(i,j;L)$ the probability for piece $(i,j)$ to form 
 hadron $L$. Then we have 
 \begin{equation}
 C(i,j;L) = A(i,j;L) C(s)
 \end{equation}
 where $A(i,j;L)$ is the probability for piece $(i,j)$ to form hadron 
 with quark flavor structure of hadron $L$ and $C(s)$ is the probability 
 for piece to form the $P$ $(s = 1)$ or $V$ $(s = 2)$ meson state. In our 
 case $C(0) + C(1) =1$. Then using  the standard recipe of the quark 
 model \cite{pd}-\cite{Gw} we obtain for the OPIS of primary hadrons
 \begin{equation}
 f(L, \tilde{p}) = g(L) /\tilde{p}
 \end{equation}
 where the plateau height $g(L)$ now depends on the flavor composition 
 and spin of the meson
 \begin{equation}
 g(L) = \sum_{i,j} A(i,j;L) C(s) r_ir_j.\label{gl}
 \end{equation}
  The plateau heights of primary mesons are given in Table 1.

From (\ref{gl}) the relation between plateau heights becomes
\begin{eqnarray}
&&g(\pi):g(\rho):g(K):g(K*):g(\eta):g(\eta'):g(\omega):g(\Phi) = \\
&&1:\frac{C(1)}{C(0)}:\frac{r_s}{r}:\frac{C(1)r_s}{C(0)r}
:\frac{1}{3}(1 + \frac{2r_s^2}{r^2}):\frac{1}{3}(2 + 
\frac{r_s^2}{r^2}):\frac{C(1)}{C(0)}:\frac{C(1)r_s^2}{C(0)r^2}.\nonumber
\label{rel}
\end{eqnarray}

Thus to fix all plateau heights we need only two adjustable parameters 
$r_s/r$ and $C(1)/C(0)$. The first one may be fixed from relation of 
heights of $\rho$ and $K*$ plateau since the $P$ meson plateau heights 
are badly disguised by resonance decay.

If we choose one of the most frequently used set of parameters 
\cite{An83}-\cite{We93} 
$r_s/r = 1/3$ and $C(0) = C(1)$ ($r = 3/7, r_s = 1/7, C(0) = 1$)
we obtain for rhs of eq.(\ref{rel})
\begin{equation}
1:1:\frac{1}{3}:\frac{1}{3}:\frac{11}{27}:\frac{10}{27}:1:\frac{1}{9}.
\end{equation}
Therefore the measurement of heights of the vector meson plateau is of 
considerable interest for checking of hadronization mechanism.

The TPIS and CF of primary hadrons can be calculated the same way and for 
symmetric TPIS we obtain
\begin{eqnarray}
&&f_{2s}(L_1, L_2; \tilde{p}_1, \tilde{p}_2) = \nonumber\\
&&\sum_{i,j,k,l} A(i,k;L_1) C(s_1) A(l,j;L_2) C(s_2) 
f_{2s}(i,k;l,j;\tilde{p}_1, \tilde{p}_2) \label{f2s}\\
&&=[f_1(L_1,\tilde{p}_1)f_1(L_2,\tilde{p}_2) + 
f_1(L_1,\tilde{p}_2)f_1(L_2,\tilde{p}_1)] \nonumber \\ 
&&+ [G(L_1,L_2) - 2g(L_1)g(L_2)]\frac{\varphi(\tilde{p}_1, \tilde{p}_2)}
{\tilde{p}_1 \tilde{p}_2} - \frac{a^2}{\tilde{p}_1 
\tilde{p}_2}\Phi(L_1,L_2;\tilde{p}_1, \tilde{p}_2)\nonumber
\end{eqnarray}
where 
\begin{equation}
\Phi_{2s}(L_1,L_2;\tilde{p}_1, \tilde{p}_2) = 2g(L_1)g(L_2) 
\Phi_{2s}^{(2)} + G(L_1,L_2)\Phi_{2s}^{(1)}
\end{equation}
and
\begin{equation}
G(L_1,L_2) = C(s_1)C(s_2) \sum_{i,j,k,l}A(i,k;L_1)A(l,j; 
L_2)r_ir_k(r_j\delta_{lk} + r_l\delta_{jl}).
\end{equation}

Thus from (\ref{f2s}) we obtain for CF of primary particles the expression
\begin{equation}
C_2(L_1,L_2;\tilde{p}_1, \tilde{p}_2) = S(L_1,L_2)\varphi(\tilde{p}_1, 
\tilde{p}_2) - a^2[\Phi_{2s}^{(2)} + (1 + S(L_1,L_2))\Phi_{2s}^{(1)}]
\end{equation}
where
\begin{equation}
 S(L_1,L_2) = G(L_1,L_2)/2g(L_1)g(L_2) - 1
\end{equation}

Apart from the oscillation contribution term all primary particle 
correlations are governed by the same function of particle rapidity 
difference, i.e. up to $\sim a^2$ corrections the all primary two particle 
CF have the same form and all dependence of CF on the type of 
particles is factorized in the scaling factor $ S(L_1,L_2)$.

It is quite obvious that the tube oscillation term is significant only 
when we detect whole nonet of particles. This is just the above 
considered case of detection of pieces irrespective to its flavor quantum 
numbers. This means that the role of tube surface oscillations may be 
studied in full nonet of vector particle correlations (since the hadron 
resonance decay badly disguise the pseudoscalar particle CF).

For individual particles the CF are strongly dominated by nonoscillation 
term. Next we consider this dominant part of CF for $P$ and $V$ 
particles. We have
\begin{equation}
C_2(L_1,L_2;\tilde{p}_1, \tilde{p}_2) = S(L_1,L_2)\varphi(\tilde{p}_1, 
\tilde{p}_2) 
\end{equation}
(The factors $S(L_1,L_2)$ are given in Table 2). 

Let us begin with CF that are independent of the particle spin. The first 
example is that both particles belong to the highest or lowest $I_3$ 
in isomultiplet ($\pi^{+} \pi^{+}, \pi^{-} K^{0}$ etc).This is just the 
same situation as we have seen above for correlation of two pieces with 
same charges. The $S(L_1,L_2) = - 1$ and the function $\varphi(\tilde{p}_1, 
\tilde{p}_2) $ can be measured directly using $K^{+*} K^{+*}, K^{-*} 
K^{-*}$ CF. It is slightly more difficult to measure the function 
$\varphi(\tilde{p}_1,\tilde{p}_2) $ using the $\rho^+ \rho^+$ or $\rho^-
\rho^-$ CF because the of distortion rising from resonance decay. 

For nonstrange particles with opposite charges we have
\begin{equation}
S(\pi^+ \pi^+) = S(\pi^+\rho^-) = S(\pi^-\rho^+) = S(\rho^+\rho^-)
=\frac{1 - r}{r}
\end{equation}
i.e. in this case the CF coincide with CF of oppositely charged pieces.
Then the measure of the relations $C_2(\rho^+\rho^-)/C_2(\rho^+\rho^+)$ 
and $C_2(\rho^-K^{*+})/C_2(\rho^+\rho^+)$ or 
 $C_2(\rho^+K^{*-})/C_2(\rho^+\rho^+)$ defines the parameters $r$ and 
 $r_s$ respectively. Thus the measurement of the CF of the vector hadrons 
 especially the CF of $\rho$ and $K^*$ mesons is highly desirable despite 
 of difficulties of data processing  and necessity to take into account
 the corrections due the resonance decay.
 
 \section{Conclusions}
 \bigskip
 
  We have seen above that the  OPIS of primary hadrons for small rapidities
 have the universal plateau form $g(L)/\tilde{p}$ where $g(L)$ is the 
 plateau height and $\tilde{p}$ is the light-cone momentum of hadron,
 where  $g(L)$ depends on the flavor composition as well as on the 
 spin of the hadron.The tube surface oscillations contribute only small 
 correction term into $g$ proportional to oscillation amplitude squared.
 
 The TPIS and CF depend on the particle detection mode. When the whole 
 multiplet of hadrons is detected irrespective of particle inner 
 quantum numbers the CF contains only the tube oscillation contribution
 term. Otherwise the oscillations give only a small correction term. The 
 main (nonoscillation) term  of CF of fixed particles has the form 
 $S(L_1,L_2) \varphi(\tilde{p}_1,\tilde{p}_2) $ where all dependence 
 on the particle type factorizes in $S$  and 
 $\varphi(\tilde{p}_1,\tilde{p}_2) $  is a function of the difference of
 particle rapidities only. Thus all primary CF have the same 
 rapidity dependence  that is scaled by $S(L_1,L_2)$ factor.
 
 The parameters of the model can be unambiguously fixed by measuring the 
 one and two particle IS of vector mesons. Such  experimental
  investigations are highly 
 desirable because of large disguise of IS of pseudoscalar particles 
 (pions and kaons) by decay of hadron resonances. We consider the 
 influence of the resonance decay on IS of pions and kaons in the next 
 paper.

\section{Acnowlegments}

Part of reported results has been previously obtained in unpublished 
article of E.G.Gurvich and myself \cite{Gg90b}. I am very grateful to Dr. 
E.Gurvich  for his collaboration at Tbilisi.

\pagebreak

\begin{table}
\caption{}
\begin{tabular}{ll}
$g(\pi) = C(0) r^2$,      &                 $g(\rho) =  C(1) r^2$,\\
$g(K) = g(\bar{K}) = C(0) rr_s$, &   $g(K^*) = g(\bar{K}^*) = C(1) 
rr_s$,\\ 
$g(\eta) = C(0)(r^2 + 2r_s^2)/3$,  &       $ g(\omega) =  C(1) r^2$,\\
$g(\eta') = C(0)(2r^2 + r_s^2)/3$,     &     $g(\Phi) =  C(1) r_s^2$.
\end{tabular}
\end{table}

Table 2: The $S(L_1,L_2)$ - factors for $\rho$ and $K*$ mesons are equal to 
corresponding $S$-factors of $\pi$ and $K$. We have omitted the charge 
labels when $S(L_1,L_2)$ are equal for entire isomultiplet.
\begin{eqnarray*}
&&S(\pi^+\pi^+) =S(\pi^-\pi^-) = -1,\\
&&S(\pi^+K^+) =S(\pi^-K^-) =S(\pi^+\bar{K}^0)=S(\pi^-K^0)=-1\\ 
&&S(K^+K^+) =S(K^-K^-) =S(K^0K^0) =S(\bar{K}^0\bar{K}^0) = -1;\\
&&S(K^+K^0) =S(K^-\bar{K}^0) = -1;\\
&&S(\pi^+\pi^-) = \frac{1 - r}{r};\\
&&S(\pi^0\pi^0) =S(\pi^+\pi^0) =S(\pi^0\pi^-) = S(\pi \omega)
= S(\omega\omega) = \frac{1-2r}{2r}\\
&&S(\pi^+K^0) =S(\pi^-\bar{K}^0) =S(\pi^+K^-) =S(\pi^-K^+) = \frac{1-r_s}{r_s}\\
&&S(K^+\bar{K}^0) =S(K^0K^-) = S(\Phi\Phi) = \frac{1-r_s}{r_s}\\
&&S(\pi^0K^+) =S(\pi^0K^-) =S(\pi^0K^0) =S(\pi^0\bar{K}^0) 
= S(K\omega) =S(\bar{K}\omega)  = \frac{1-4r_s}{4r_s}\\
&&S(K^+K^-) =S(K^0\bar{K}^0) =\frac{1}{2}(\frac{1 - r}{r}+
\frac{1 - r_s}{r_s})\\
&&S(\pi,\eta) =-1+\frac{r}{r^2+2r_s^2}\\
&&S(\pi\Phi) = S(\omega\Phi) = -1\\
&&S(K\Phi) =S(\bar{K}\Phi) = \frac{1-2r_s}{2r_s}\\
&&S(K\eta) =S(\bar{K}\eta) =-1 + \frac{3(r+4r_s)}{2(r^2+2r_s^2)}\\
&&S(K\eta') =S(\bar{K}\eta') =-1 + \frac{r+r_s}{2(2r^2+r_s^2)}\\
&&S(\eta\eta) =-1+ \frac{r^3+8r_s^3}{2(r^2 +2r_s^2)^2}\\
&&S(\eta'\eta') =-1+ \frac{2r^3+r_s^3}{(2r^2 +r_s^2)^2}\\
&&S(\eta\eta') =-1+ \frac{r^3+2r_s^3}{(r^2 +2r_s^2)(2r^2 +r_s^2)}
\end{eqnarray*}

\newpage
{\bf FIGURE CAPTIONS}
\bigskip
1.The integration area for OPIS.

2.The two contributions to the TPIS: a)the three breaking point 
contribution; b)the four breaking point contribution.

3.The function $\varphi$ versus particle rapidity difference 
 $y=\ln(\tilde{p}_2/ \tilde{p}_1)$

\newpage
\begin{figure}
\vspace{6.0in}
\includegraphics{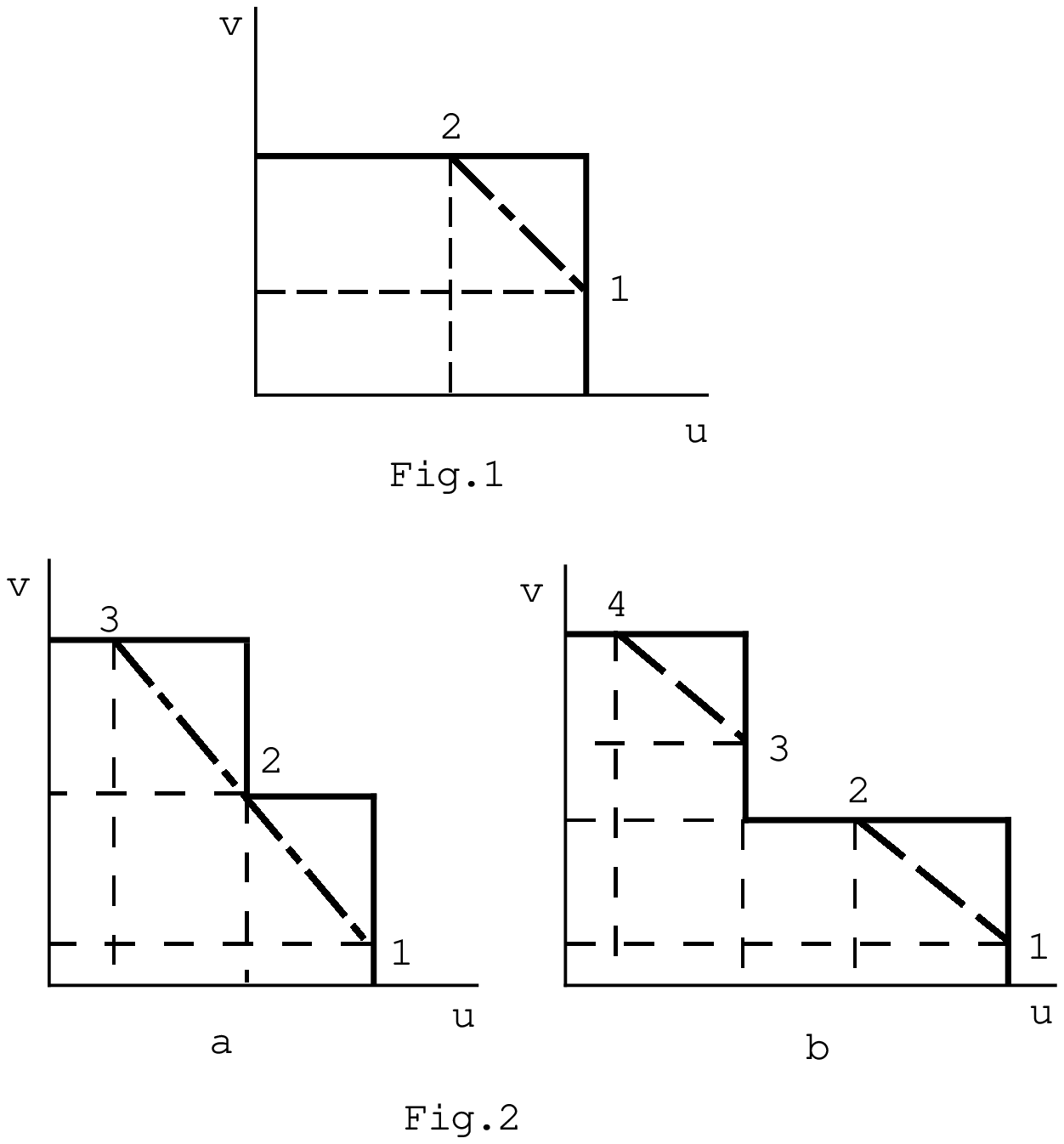}
\end{figure}

\newpage
\begin{figure}
\vspace{6.0in}
\includegraphics{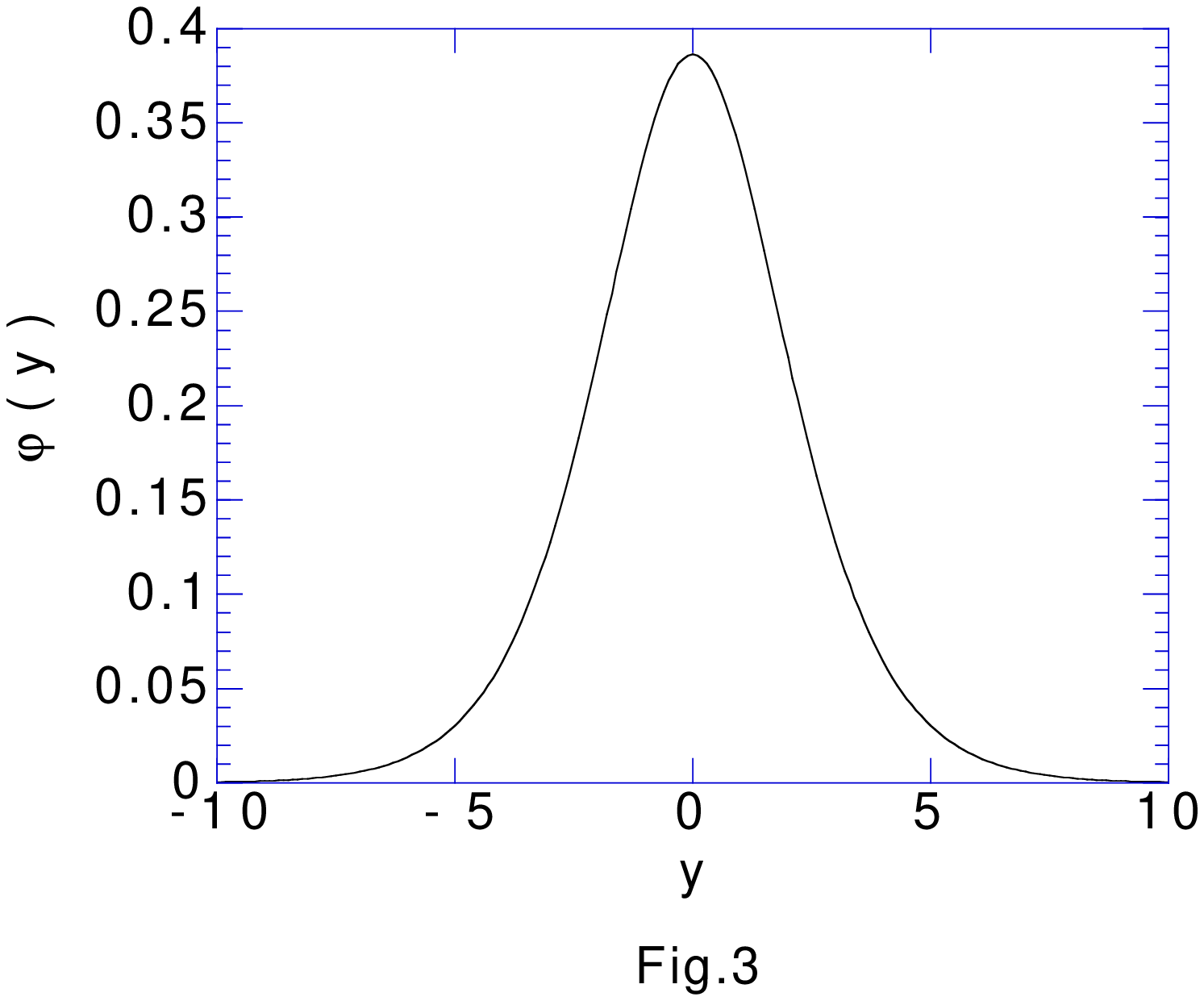}
\end{figure}

\end{document}